\documentclass[journal abbreviation]{copernicus}

\usepackage{color}
\newcommand{\refA}{ \color{black}}
\newcommand{\refB}{ \color{black}}

\begin{document}

\title{Solar stereoscopy - where are we and what developments do we require to progress?}

\author[1]{Thomas Wiegelmann}
\author[1]{Bernd Inhester}
\author[1,2]{Li Feng}
\affil[1]{Max-Planck-Institut f\"ur Sonnensystemforschung,
Max-Planck-Strasse 2, 37191 Katlenburg-Lindau, Germany}
\affil[2]{Purple Mountain Observatory, Chinese Academy of Sciences, 210008, Nanjing, China}


\runningtitle{Coronal stereoscopy}

\runningauthor{Wiegelmann et al.}

\correspondence{Wiegelmann \\ (wiegelmann@mps.mpg.de), Annales Geophysicae, Volume 27, Issue 7, 2009, pp.2925-2936}

\received{}
\pubdiscuss{} 
\revised{}
\accepted{}
\published{}


\firstpage{1}

\maketitle

\begin{abstract}
Observations from the two STEREO-spacecraft give us for the first time the possibility
to use stereoscopic methods to reconstruct the 3D solar corona. Classical stereoscopy
works best for solid objects with clear edges. Consequently an application of classical
stereoscopic methods to the faint structures visible in the optically thin coronal plasma
is by no means straight forward and several problems have to be treated
adequately: 1.)First there is the problem of identifying one dimensional structures
-e.g. active region coronal loops or polar plumes- from the two individual EUV-images
observed with STEREO/EUVI.
2.) As a next step one has the association problem to find corresponding structures in
both images. This becomes more difficult as the angle between STEREO-A and B increases.
3.) Within the reconstruction problem stereoscopic methods are used to compute the
3D-geometry of the identified structures.
Without any prior assumptions, e.g., regarding the footpoints
of coronal loops, the reconstruction problem
has  not one unique solution.
4.) One has to estimate the reconstruction error or accuracy of the reconstructed 3D-structure,
which depends on the accuracy of the identified structures in 2D, the separation
angle between the spacecraft, but also on the location, e.g., for east-west
directed coronal loops the reconstruction error is highest close to the loop top.
5.) Eventually we are not only interested in the 3D-geometry of loops or plumes,
but also in physical parameters like density, temperature, plasma flow, magnetic
field strength etc. Helpful for treating some of these problems are coronal magnetic
field models extrapolated from photospheric measurements, because observed EUV-loops
outline the magnetic field. This feature has been used for a new method dubbed
'magnetic stereoscopy'. As examples we show recent application to active region
loops.
\end{abstract}


\introduction
The Solar TErrestrial RElations Observatory (STEREO) observes for the first
time simultaneously the Sun from two vantage points \citep[see][]{kaiser:etal08}
and allows a three dimensional view of the solar corona. Within this work we
aim to review stereoscopic methods to reconstruct the 3D corona and will
concentrate mainly on structures like active region loops and polar
plumes observed with the STEREO/SECCHI instrument package
\citep[Sun Earth Connection Coronal and Heliospheric Investigation,][]{howard:etal08}.
Before the launch of STEREO on October 26, 2006 stereoscopic methods have
been applied for example to Skylab images by \citet{berton:etal85} and \citet{batchelor94},
and to SOHO data by \citep{aschwanden:etal99,aschwanden:etal00}.
In these pre-STEREO cases the authors used the solar rotation and took images
a few hours to a day apart to reconstruct the 3D structures under the assumption that
all basic features remain stationary within this time.
The stationarity assumption was somewhat relaxed by introducing the
concept of {\it dynamic stereoscopy} \citep{aschwanden:etal99}, which
uses the a priori information of a
{\refA coplanar loop shape and a corresponding fitting procedure.
In this approach the coronal magnetic field -but not necessarily the
plasma-  is considered to be
quasi-stationary. The method takes advantage of the near-parallelity of
adjacent magnetic field lines, even if the loop plasma is heated and cools down on
much faster time scales  than the time interval of stereoscopic
correlation.}
 Such limitations are not necessary anymore after
the launch of STEREO about two and a half year ago and within this
paper we would like to give a review on what has been done in
solar stereoscopy so far, which developments are currently under
consideration, and planned for the future.

The key question is how we can derive the 3D geometry and physical structure
of the solar corona  from images observed with the two STEREO-spacecraft?
In (section \ref{sec2}) we describe {\it a step by step guide}, which contains
the identification of curvi-linear structures from coronal EUV-images in (section
\ref{sec2.1}), the association of the identified structures in both images
from different vantage viewpoints in (section \ref{sec2.2}), the geometric 3D stereoscopy
(section  \ref{sec2.3}) and estimation of the 3D-reconstruction error in
(section  \ref{sec2.4}). After these steps one has obtained the 3D geometry
of, e.g., active region loops or polar plumes and in (section \ref{sec2.5})
we outline how physical quantities like temperature and density can be
found. An interesting question, which we address in (section \ref{sec3}),
is how well do stereoscopic reconstructed plasma loops agree with coronal
magnetic field models? Due to the high conductivity of the coronal plasma
the magnetic field is outlined by the radiating plasma and in principle
one has two independent data sources about the 3D geometry of coronal loops,
namely stereoscopy and magnetic field extrapolations from photospheric
measurements. In (section \ref{sec4}) we address how in future stereoscopic,
tomographic and self-consistent modelling approaches could be combined
and diminish weaknesses of the individual approaches.
\section{Step by step guide to stereoscopy}
\label{sec2}
\subsection{Extraction of curvi-linear objects from EUV-images}
\label{sec2.1}
\begin{figure}
\vspace*{2mm}
\begin{center}
\includegraphics[width=8.3cm]{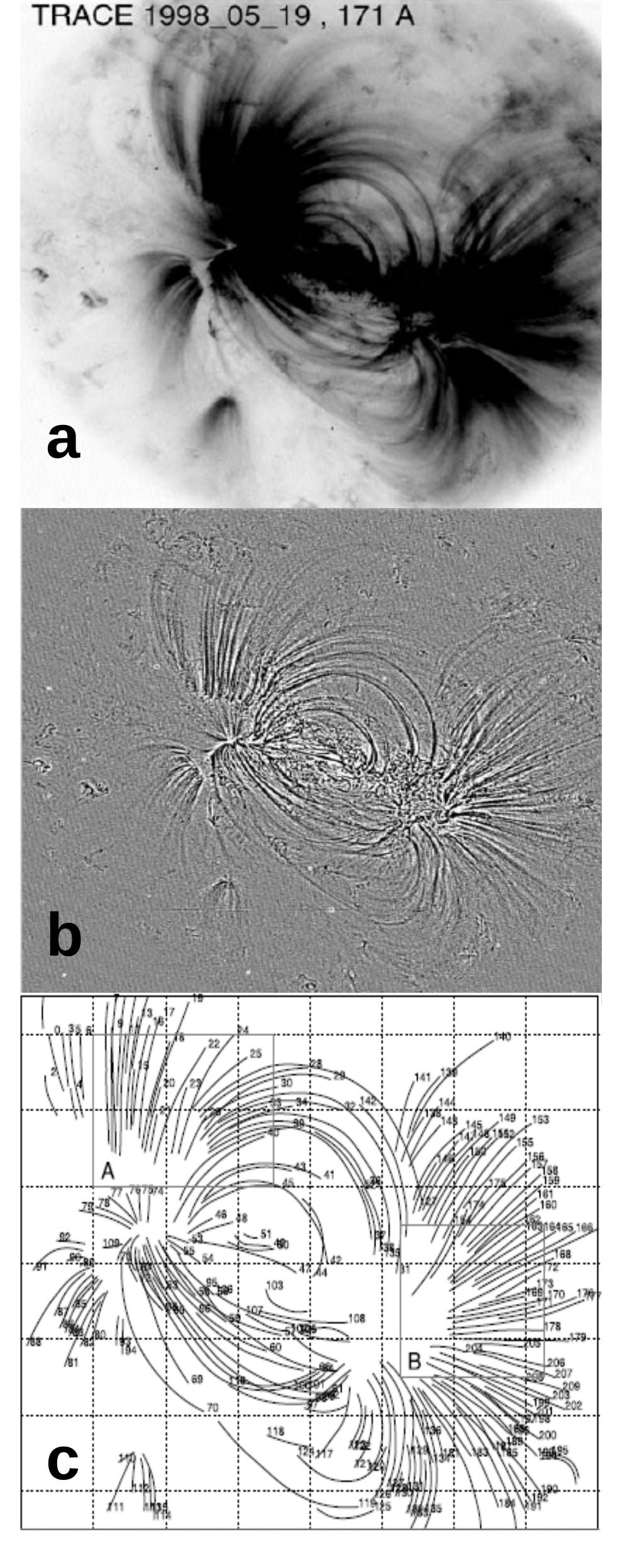}
\end{center}
\caption{Panel a) shows an original 171 \AA $ $ TRACE image of 19 May 1998,
panel b) a corresponding high-pass-filtered image, where a
$7 \times 7$-boxcar smoothed image was subtracted from the original.
The image has been used to compare the output of
five feature recognition codes with the result of 210 manually traced loop
shown in panel c).
[Original figures are from \citet{aschwanden:etal08} figure 1 and 2].}
\label{fig1}
\end{figure}
A first step for stereoscopy is to extract curve-like structures (projections of
coronal loops) from observed EUV-images such as from the TRACE-image shown in figure
\ref{fig1} a). A principal problem of identifying loops is that  the
solar corona is optically thin and the loops are faint. Visible loops are often
a superposition of multiple individual loops \citep{schrijver:etal99,schrijver:etal04}.
Figure \ref{fig1} b) shows the TRACE-image after a $7 \times 7$-boxcar smoothed image
was removed from the original, which enhances the contrast.
\citet{aschwanden:etal08} manually traced 210 loops as
shown in panel c). While hand-tracing might be suitable to investigate a few
individual cases, this is not appropriate to study large data sets and time sequences.
Several automatic feature recognition methods have been developed. The
example shown in figure \ref{fig1} has been used to compare and evaluate
five automated loop segmentation methods.
The output of the five codes has been compared
with the hand-traced loops from  \ref{fig1} c). This comparison
revealed large differences. The codes identified between $76$ and $347$ loops.
Among the longer and more significant loops, the various codes identified between
$19 \%$ and $59 \%$ of the corresponding 154 hand traced loops above this limit.
Some codes wrongly identified noise as spurious short loops. Obviously the
state-of-the-art of automatic feature recognition techniques is not satisfactory.
The individual codes have control parameters, adapted to the image signal to
noise ratio, resolution and also to the type of objects it aims to extract.
{\refA A prime problem is that time-varying background loops
and moss prevent an uncontaminated separation of loop and background.
This makes it difficult to trace loop tops and footpoints.
The problem becomes more complex for the comparison of EUV-images taken at
different wavelengths and temperatures because the background is
different in each filter.}
In the current stage the automatic
feature recognition tools are already useful, but some human interaction
is usually necessary and the codes can be considered as semi-automatic.
The automated segmentation of polar coronal plumes, which we had during the
recent solar minimum ample opportunity to observe
{\refB \citep[see][for details]{feng:etal09}}, is more advanced. Even
though they are fainter in intensity than loops, their shape is more restricted
so that wavelet and the Hough transform can be employed for their detection.
{\refB \citet{llebaria:etal02} applied the  Hough transform to a time series
(66 hours) of SOHO/LASCO-C2 observations to generate time intensity diagrams
of polar plumes. Such automated plume detection can be applied also to
STEREO-images from two viewpoints (de Patoul et al., in preparation).
An alternative to stereoscopy is to study polar plumes with tomographic
methods \citep[see][for an application to SOHO/EIT images under the model assumption
of 'polar plumes as stationary objects whose intensity varies homogeneously with time.']{barbey:etal08}.
Two STEREO-viewpoints are expected to be most useful for the tomography of plumes
when the separation angle is about $60^\circ$.}

\subsection{Association of objects in both images}
\label{sec2.2}
\begin{figure*}
\vspace*{2mm}
\setlength{\unitlength}{1.0cm}
\begin{picture}(15,15)
\put(-0.7,3.0){\includegraphics[width=14.3cm]{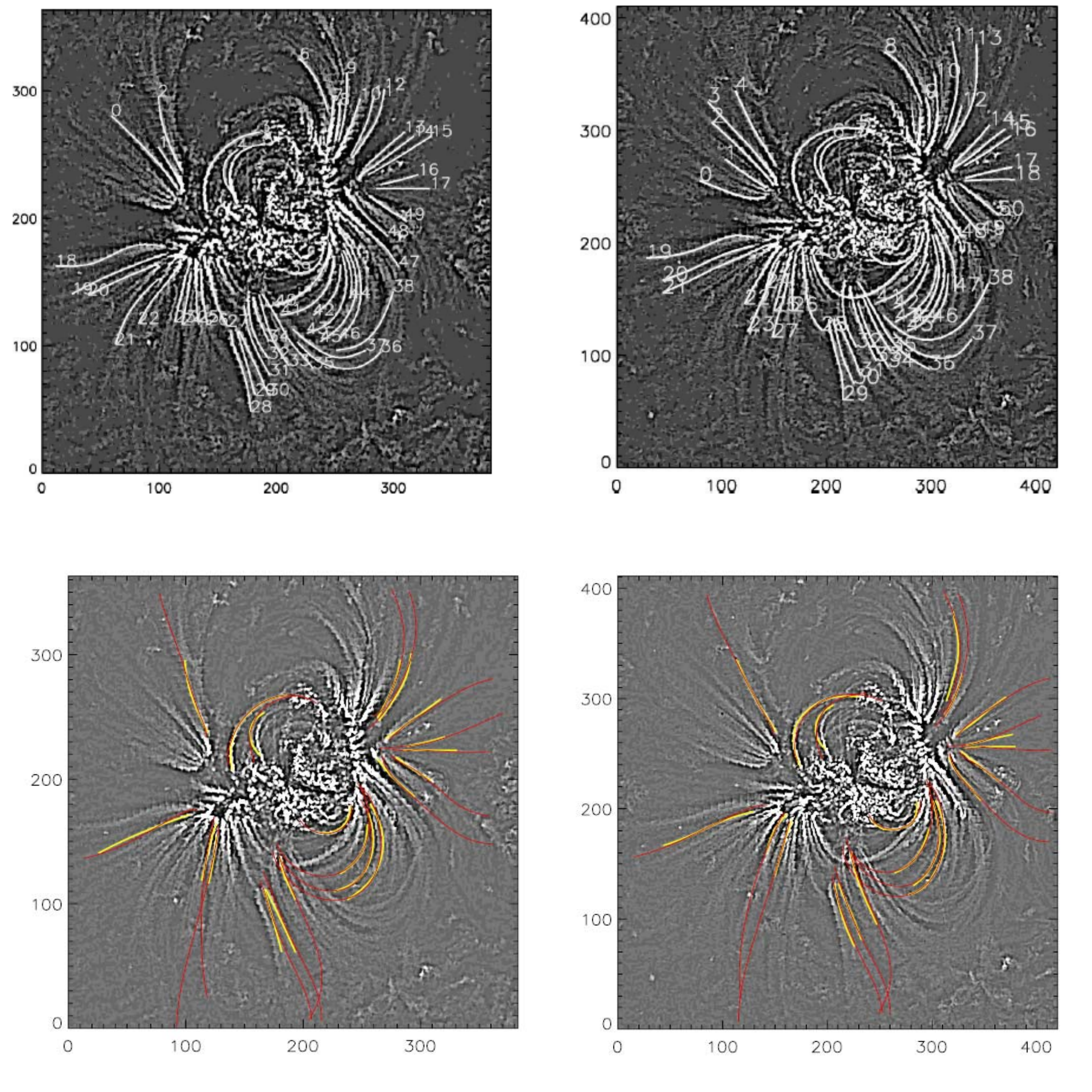}}
\end{picture}
\caption{Top: Contrast enhanced EUVI images from STEREO-B (left) and A (right) of the NOAA AR 0960 on
8 June 2007. The individual loops (enumerated white curves) have been extracted by
a semi-automatic feature recognition tool as described in \citet{inhester:etal08}.
Please note that equal numbers do not imply a correspondence across the
images. Bottom: Some selected coronal loops (yellow) with their best fit magnetic
field line (red).
The separation angle between the spacecraft was $12^\circ$.
[The top figures have been published originally in \citet{feng:etal07} figure 1.]}
\label{fig2}
\end{figure*}
For faint objects like coronal EUV-loops it is not trivial to find associated
structures in images taken from different vantage points as shown in figure \ref{fig2}.
\citet{rodriguez:etal09} developed a correlation tracking method which
automatically matches pixels in both images, which worked well for small
separation angles between spacecraft but it becomes difficult and ambiguous
for if the separation angle of the spacecraft exceeds about $15^\circ$.
In particular for large separation angles some structures might be visible
in one image, but not in the other. \citet{aschwanden:etal08a} applied a
forward projection with an assumed height range of $h=0 \dots 0.1 R_s$, which was sufficient
to find correspondence for 30 traced loop, when the separation angle
between STEREO-A and B was only $7^\circ$. The correspondence problem
becomes more difficult to solve for larger separation angles.

Other possibilities
are the use of a priori assumptions of the coronal structures, e.g., fitting
to a semi-circular loop model \citep{aschwanden:etal99}, or loop curvature
constraints \citep{aschwanden05}. As an alternative one can use the fact that
the emitting EUV-radiation outlines magnetic field lines
due to the high conductivity of the coronal plasma.
Consequently magnetic field lines should
provide a reasonable proxy for coronal plasma loops.
\citet{wiegelmann:etal02} used the stereoscopic reconstructed loops
from \citep{aschwanden:etal99} and photospheric magnetograms from SOHO/MDI
to fit the optimum parameter $\alpha$ within the linear force-free field
approximation. Projections of extrapolated 3D magnetic field lines under
different model assumptions have been compared with coronal images
in \citep{gary:etal99,carcedo:etal03,regnier:etal04,wiegelmann:etal05b}.
The method has been extended for a STEREO pair of
EUV-images in combination with linear and nonlinear force-free field models
in \cite{wiegelmann:etal06c} and was dubbed {\it magnetic stereoscopy}.
The idea of magnetic stereoscopy is that a number of 3D field line proxies are
projected onto the EUV-images and compared with the corresponding extracted
curve-like structures (see section \ref{sec2.1}). Loops in both images
which have a minimum distance to the projection of a 3D magnetic field
line are very likely related to each other. The field line proxies where
generated from extrapolation models but any other method to produce
parameterized meaningful 3D curves  would work as well.
{\refA The extrapolation models are here just a convenient means to generate
3D curves the observed loops can be compared with.}
The method has been applied
to TRACE-data (taken a day apart) by \citet{feng:etal07} and to
STEREO/SECCHI by \citet{feng:etal07a}. In both cases the magnetic field
has been computed with the linear force free method developed by
\citet{seehafer78} from SOHO/MDI. Force-free fields are characterized
by $\nabla \times B = \alpha B$ and $B \cdot \nabla \alpha=0$, where $B$
is the magnetic field and $\alpha$ is zero for potential fields and constant
in the entire space for linear force-free models.
For more sophisticated non-linear force-free field models, $\alpha$ is
constant along field lines but may vary on different field lines.
An advantage of using coronal magnetic
field models is that they generate meaningful and physics-based
3D curves. Note, however, that a set of field lines from a linear
force-free model constructed with a different $\alpha$ does not
constitute a physically consistent magnetic field model.
Disadvantages are that one needs additional
observations from ground-based or space-borne magnetographs
(a third eye, e.g., SOHO or in future SDO) and that magnetic modelling and
stereoscopy are not independent from each other, which is helpful for
evaluating the consistency of both methods (see also discussion in
section \ref{sec3}).
\subsection{Geometric Stereoscopy}
\label{sec2.3}
\begin{figure}
\vspace*{2mm}
\begin{center}
\includegraphics[width=8.3cm]{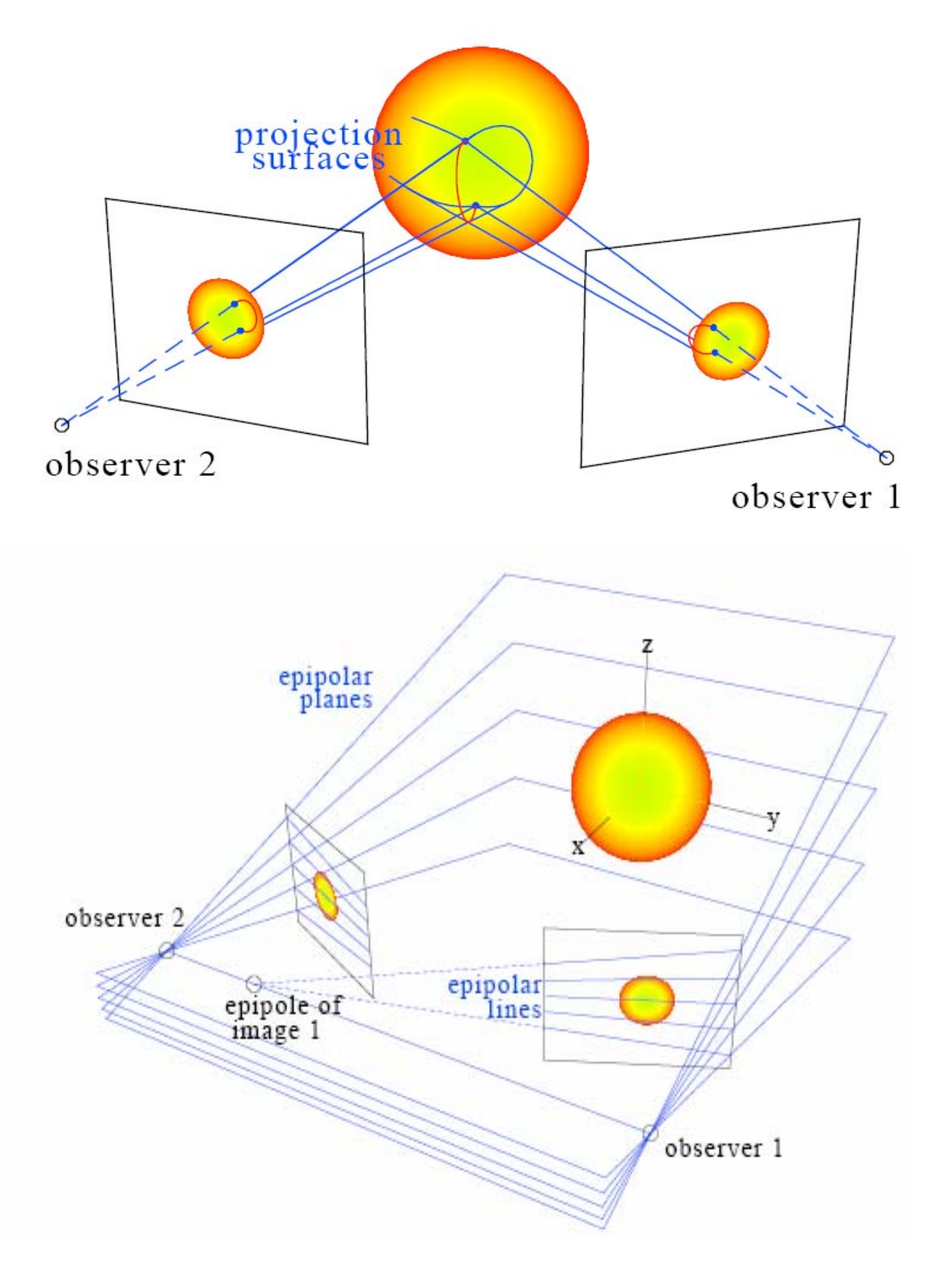}
\end{center}
\caption{{\refB Top panel: Back projection to reconstruct curve-like
objects, e.g., coronal loops, from two images.
Bottom panel: Epipolar planes and the corresponding epipolar lines in
two STEREO-images.}
[Top and bottom panel of this figure have been originally published
in \cite{inhester06} figure 1 and 2, respectively.]}
\label{fig3}
\end{figure}
After having solved the feature extraction and association problem, the 3D
reconstruction is in principle only a geometric problem as indicated in
figure \ref{fig3} top panel. For stereoscopic reconstruction,
it is helpful to introduce a suitable coordinate
system - called {\it epipolar geometry} - which reduces the original 3D reconstruction
problem to a number of 2D problems.
The positions of the two STEREO-spacecraft and any object point define a plane,
called {\it epipolar plane}, which serve as a natural coordinate system for
the stereoscopic reconstruction (see figure \ref{fig3} bottom panel).
The projection of the epipolar planes onto the two STEREO image planes are
called {\it epipolar lines}. Per definition all epipolar lines in image-A converge
at one point (marked as epipole of image 1 in figure \ref{fig3} bottom panel)
and vice versa. In practice the epipolar lines appear almost (but of course not strictly)
parallel in images from the STEREO-EUVI \citep[see][for quantitative estimations]{inhester06}.
As the STEREO-spacecraft are close
to the ecliptic the epipolar lines intersect with the rotation axis of the
Sun (indicated as the z-axis in figure \ref{fig3} bottom panel) and the angle with this
axis can be used to label the epipolar lines. All points in space, except points on
the stereo base line (the line which connects both spacecraft), are lying on
a uniquely defined epipolar line. This is a helpful constraint as all points,
visible on a specific epipolar line in image-A must lie also on the same
epipolar line in image-B. In particular, for large separation angles, there
is  no guaranty that a faint coronal structure identified at observer-A
will be also visible at observer-B.
 This so called {\it epipolar constraint}
\citep{inhester06} is very helpful because it reduces the 3D-stereoscopic reconstruction
to calculating the position of each object on it's uniquely defined 2D epipolar plane.
The difference of the positions of an object projected along the epipolar lines in
the two STEREO-images yields the depth information or in case of coronal loops the
height in the corona.
A simple way to tie-point associated loop curves in image A and B is to label
along each curve the intersections with given epipolar lines and to reconstruct
the intersection points with the same epipoles line label. This, of course,
requires that both curves cover more or less the same epipolar range,
a criterion which can be used to confirm a correct association.

\subsection{Estimating the reconstruction error in 3D}
\label{sec2.4}
\begin{figure}
\vspace*{2mm}
\begin{center}
\includegraphics[width=8.3cm]{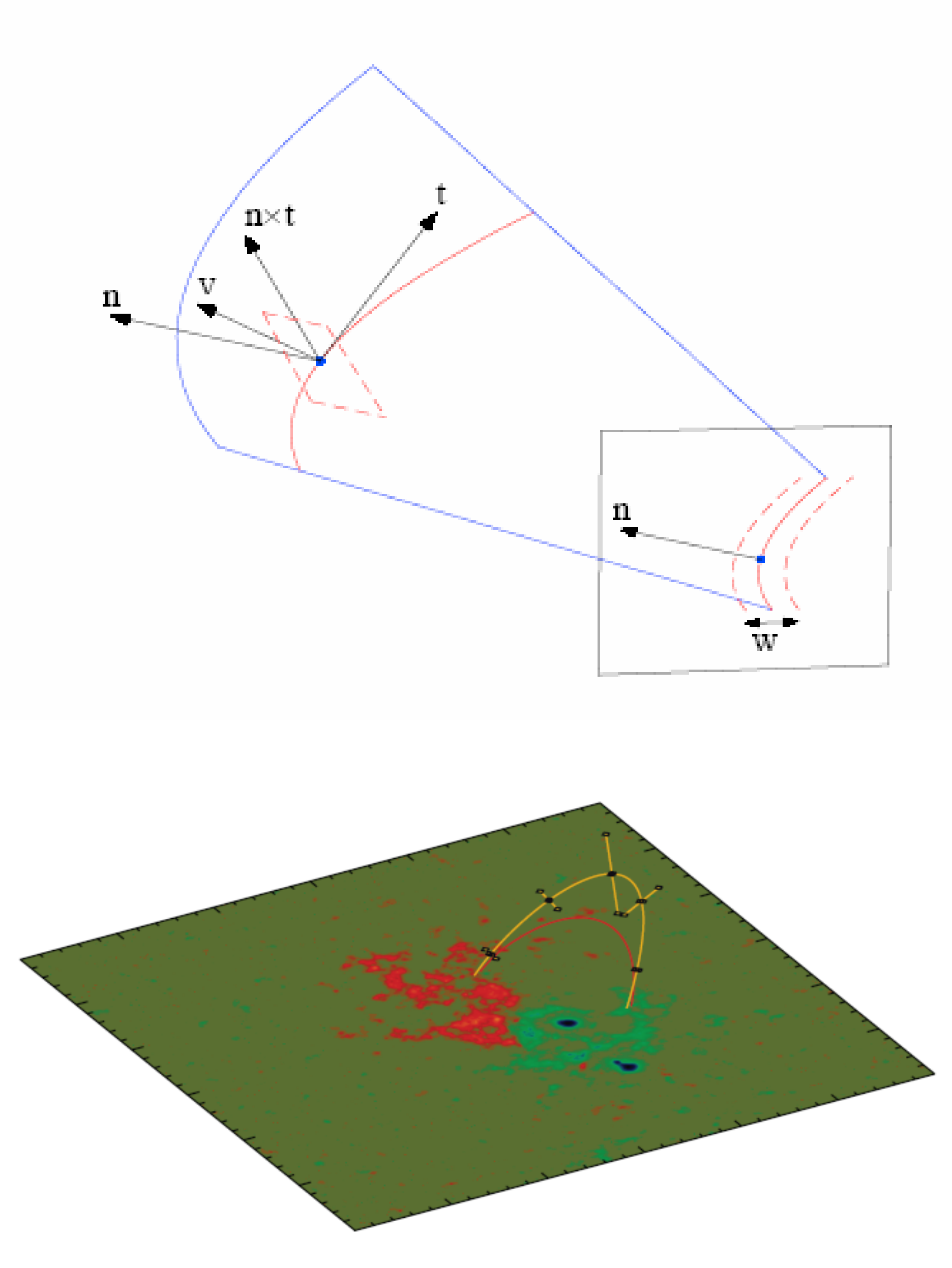}
\end{center}
\caption{Top: How does the uncertainty $w$ of the projected
loop in both EUV-images affect the error-trapezoid of the
reconstructed 3D-loop? (see text)
[Original publishes in \citep{inhester06} figure 10],
Bottom: Stereoscopic reconstructed 3D loop (yellow) and
best fit linear force free coronal magnetic field line
extrapolated from SOHO/MDI magnetograms.
[Original published in \citep{feng:etal07a} figure 5.]}
\label{fig4}
\end{figure}
Features in the two STEREO-images can be identified of course only within
a certain error margin as indicated by $w$ in figure \ref{fig4} top panel.
Possible error sources are the finite resolution of the instrument as
well as uncertainties occurring due to extracting features from the
EUV-images. The question is how do these uncertainties in the 2D-images
affect the reconstruction error of the 3D coronal loop? As a consequence
of the finite resolution $w$ of the loop projection, the true 3D
coordinate of a loop point lies nearby in a plane through the
reconstructed point which is spanned by the
local normals $n_i$ of the two projection surfaces.
$t=(n_1 \times n_2)/|n_1 \times n_2|$
is the local curve tangent. Decomposing the uncertainty vector
into its components $n_1$ and $n_2$ yields an
error trapezoid of the positional uncertainty
has the axes $w/\left( 2 \cos \left(\alpha/2 \right) \right)$
and $w/ \left(2 \sin \left(\alpha/2 \right) \right)$,
where $\alpha$ is the angle between the projection surface normals
\citep[see][for a mathematical derivation and more details]{inhester06}.
For small $\alpha$ the depth error along the mean view direction
of the two spacecraft may be considerable:
$1/ \left( 2 \sin(\alpha/2) \right)$ exceeds $5$ for
$\alpha < 10^\circ$.
$\alpha$ is limited from above by the angle between the two
STEREO-spacecraft. This upper limit is reached when the loop
intersects an epipolar line normally. This case allows the
most accurate 3D reconstruction for a given separation angle between
spacecraft. For features parallel to epipolar lines $\alpha$
is zero and the 3D reconstruction error becomes infinite. As epipolar
lines are almost horizontal in the images this means that
the highest 3D reconstruction error occurs at the top of
east-west loops. As a consequence, there are large error
bars in figure \ref{fig4}
bottom panel, which shown a stereoscopic reconstruction
of one loop from \citep{feng:etal07a} for a separation angle
of $12^\circ$ between spacecraft. A small (large) separation angle
between spacecraft makes it easy (difficult) to associate
related structures in both images, but for the 3D reconstruction
error it is the other way around, a small separation angle leads
to a large error in 3D.

{\refA An alternative error estimate has been given by
\citet{aschwanden:etal08a} for the uncertainty in height.
Both error estimates show a similar behaviour, however: The smallest error is
obtained if the loop tangent is normal to the epipolar planes and
the error becomes infinite if the loop tangent becomes parallel to the epipolar
plane.}

%
%
\begin{figure}
\vspace*{2mm}
\begin{center}
\includegraphics[width=8.3cm]{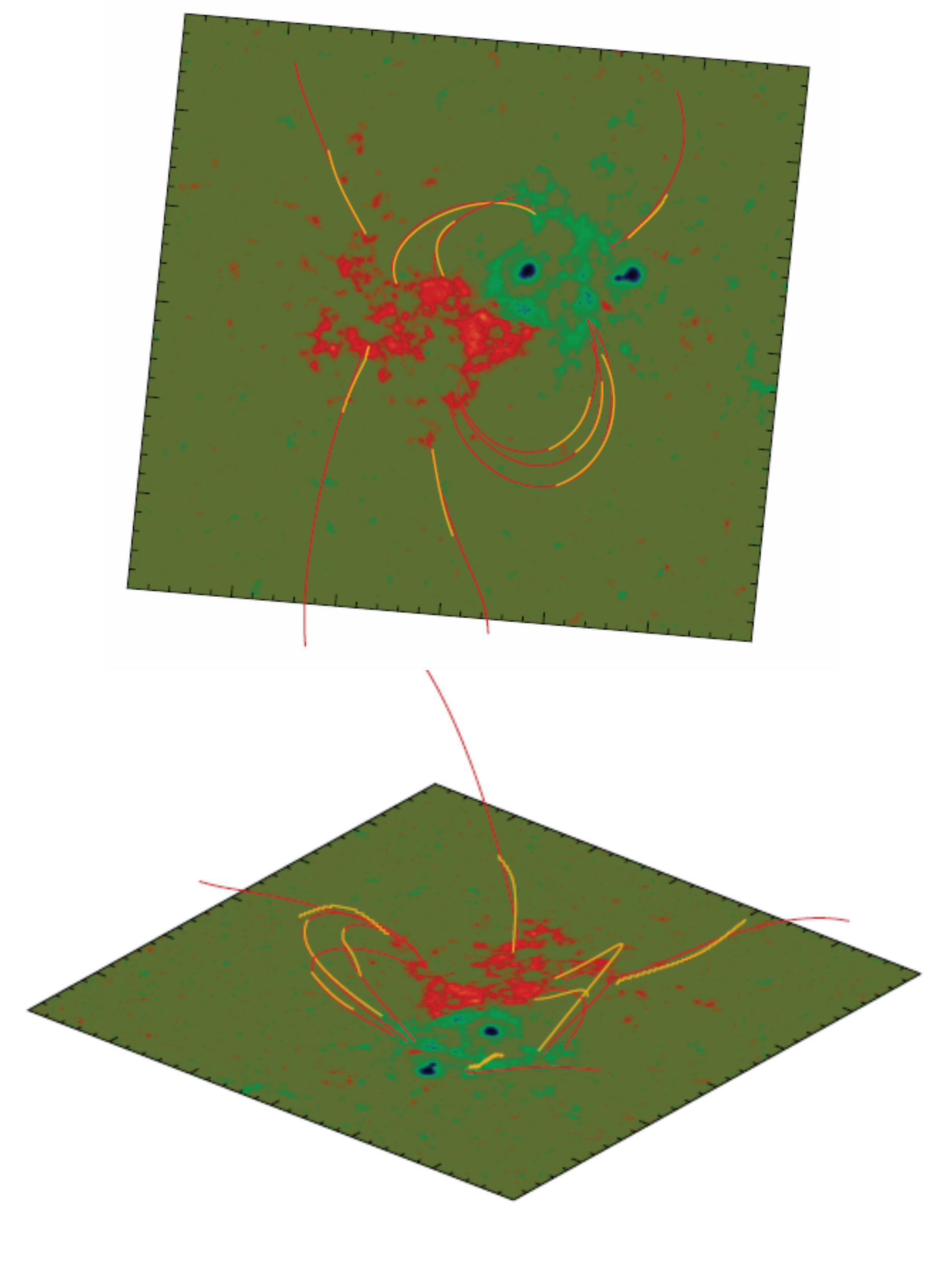}
\end{center}
\caption{Stereoscopic reconstructed 3D loops and sections of loops
(yellow) and best fit linear force-free field lines (red) from different
viewpoints (top panel: view from STEREO-A, bottom panel: Northeast of
the Active Region).
[Original published in \citep{feng:etal07a}, figure 3 and 4.]}
\label{fig6}
\end{figure}

Another problem in finding a unique solution for the 3D loop is
a possible reconstruction ambiguity.
{\refA This is a problem which can theoretically
occur even for a pair of correctly identified
loop projections, if the footpoints are wrongly associated in the image pair.
In EUV images, the footpoints of loops are sometimes difficult to locate.
They may be drowned in bundles of other loops or near-surface EUV moss.
This problem does, however, disappear if one
can identify the footpoints of loops
and requires them to be located near the solar surface. It is,
however, not always possible to identify the loop footpoints and
in this case one has two possible candidates for the true 3D-solution.}
The true and the false reconstruction intersect
at a point, where the projected segment is parallel to an epipolar
line, e.g., the top of east-west coronal loops where the above error
estimate formally diverges.

How can ambiguities and errors be limited? One possibility would be to use additional
EUV-images from a third viewpoint, e.g., SOHO/EIT or in future SDO/AIA.
This possibility has (to our knowledge) not been tried out yet.
A potential problem might be the different resolution of the STEREO/SECCHI-EUVI
and the SOHO/EUV instruments. The error in line-tying is also reduced by making
use of the fact that the reconstructed loops should resemble field lines and hence
should be smooth. Smoothing and/or spline fitting of the tie point reconstruction
should in general reduce the reconstruction error.

Another possibility, which has been already
shown to be useful for the association problem (section \ref{sec2.2}), is
to use coronal magnetic field models. This possibility, called {\it magnetic stereoscopy}
was first tested with a model active region  in \citep{wiegelmann:etal06c}. Different
magnetic field models, potential, linear and non-linear force-free have been
used and it was shown that even the use of a poor field model (potential fields)
was sufficient to resolve the reconstruction ambiguity. The extrapolated magnetic
field lines provide already a proxy for the 3D plasma loop and if ambiguous
solutions occur, the solution closer to this proxy-loop is chosen. Figure
\ref{fig4} bottom panel and \ref{fig6} shows a linear force-free field line (extrapolated
from SOHO/MDI) in red, which
has been used for this aim in \citep{feng:etal07a}. Extrapolated field lines
might also serve as a reasonable approximation of the plasma loop in regions
with a large reconstruction error, e.g., the loop top or if only parts of
the loop are visible in the EUV-images.
\subsection{Derive physical quantities}
\label{sec2.5}
\begin{figure}
\vspace*{2mm}
\begin{center}
\includegraphics[width=8.3cm]{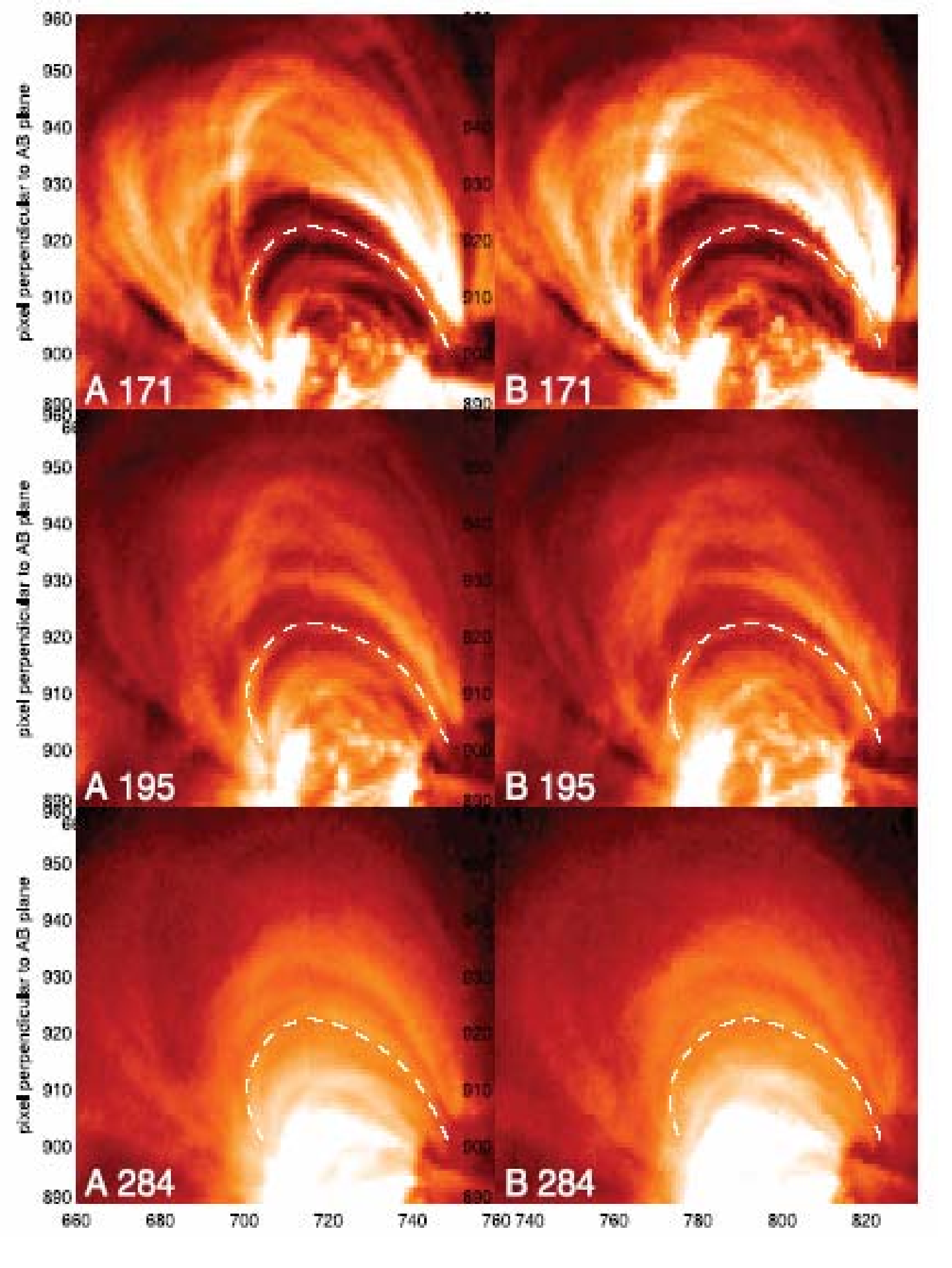}
\end{center}
\caption{EUVI-images at different wavelengths from STEREO-A and B in
the left and right panels, respectively. The dashes white line
shows a projections of a 3D loop as reconstructed
in \citep{aschwanden:etal08a}.
[Original published in \citep{aschwanden:etal08b} figure 1] }
\label{fig7}
\end{figure}
\begin{figure}
\begin{center}
\includegraphics[width=8.3cm]{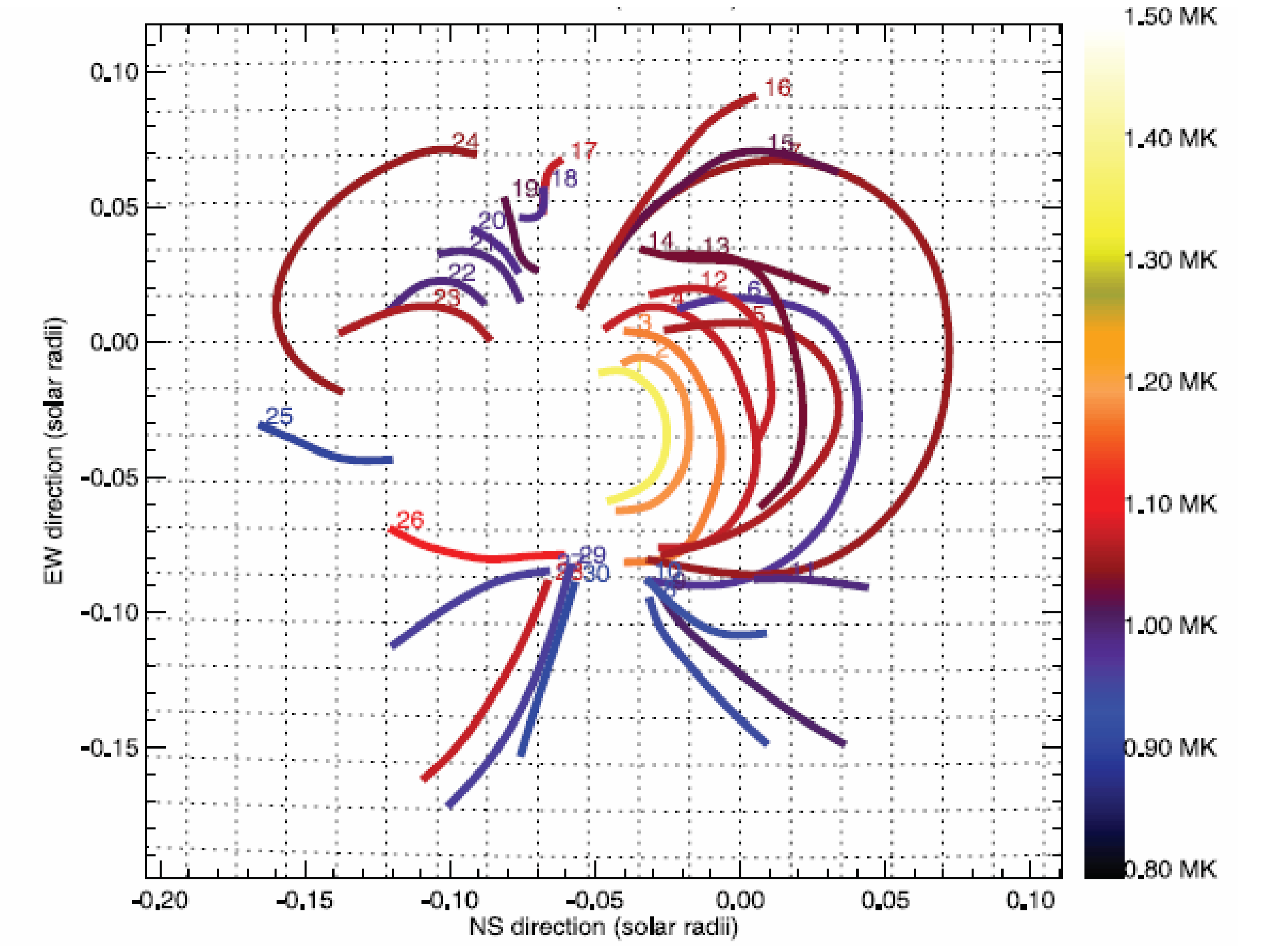}
\end{center}
\caption{Temperature map of 30 loops. The loops are isothermal within
the error estimation and the hottest loops are also the smallest.
[Original published in \citep{aschwanden:etal08b} as part of figure 9].}
\label{fig8}
\end{figure}
The geometry of the 3D coronal structures, as seen from different viewpoints in
figure \ref{fig6} in yellow provide already useful information. \citet{feng:etal07a}
found that most of the reconstructed 3D loops cannot be approximated
by planar  curve segments and that most of the loops are not circular.
This was already known from field modelling.
E.g. meaningful loop emissions per unit loop length can only be
derived from EUV images, if the angle between the viewing direction and
the loop tangent is properly taken into account \citep{aschwanden:etal08b}.

In the following we will concentrate on the computation of physical
plasma parameters, in particular the electron temperature and
density along the reconstructed 3D loops. Figure \ref{fig7} shows
the projection of a 3D loop as reconstructed in \citep{aschwanden:etal08a}
onto STEREO/SECCHI images at different wavelengths taken from STEREO-A
and B in the left and right panels, respectively. As explained in detail
in \citep{aschwanden:etal08b} the EUVI-images can be used to obtain
the electron temperature and density, independent of each other
for the EUVI-images taken from both STEREO-viewpoints, which have
been separated by $7^\circ$ for this study. As only a fraction of about
$10 \%$ of the EUV-radiation is coming from the loops, one has to remove
first the background separately in each wavelength, which is a tricky
busyness and requires model assumptions as explained in detail in
\citep{aschwanden:etal08b}. After the well known temperature
response functions (defined for each wavelength), as calculated with
the CHIANTI-code, are used to obtain physical quantities along
the loops. For this aim a local loop-aligned coordinate system is
introduced and the differential emission measure (DEM) is
constrained for each loop position with 3 temperature filters
at various positions along the loop with a Gaussian function.
The DEM is then fitted with the help of the EUVI response functions
in three wavelength, which provides approximations for the temperature,
the Gaussian half width of DEM and the peak emission measure at the
chosen positions along the loop
\citep[see][equations 1-4 for details]{aschwanden:etal08b}.  Figure \ref{fig8}
shows the temperature for 30 plasma loops. The loop-temperatures
computed with this technique where found isothermal along the loop
(within an error margin of about $20 \%$ caused mainly by
background substraction uncertainties).
The temperature scales with loop length with
the shortest loops are also the hottest ones. For a known
loop width (as shown in figure \ref{fig4}, upper panel) the
three parameters of the DEM-model (temperature,
Gaussian half width, peak emission measure) can be converted into
electron densities
\citep[see][section 3.4, equations 5-10 for details]{aschwanden:etal08b}.
With the stereoscopic reconstructed 3D-loop projected onto images
from STEREO-A and B the temperature and density can be calculated
independently from both spacecraft. There have been some systematic
differences regarding the results from both spacecraft, e.g., the
average loop temperature computed from STEREO-B was somewhat higher as
the one from STEREO-A, but the discrepancies have been only
a few percent for density and temperature estimations. It is still
to be investigated how consistent the estimations from both spacecraft
are for larger separation angles. Coronal stereoscopy provides us for the
first time with the  3D-loop length, temperature and density. These quantities
can be used to test RTV-scaling laws \citep{rosner:etal78} and to go a
further step towards a self-consistent modelling of the solar corona,
which will be discussed in section \ref{sec4}.
\section{Stereoscopy and coronal modelling}
\label{sec3}
\begin{figure*}
\vspace*{2mm}
\begin{center}
\includegraphics[width=15cm]{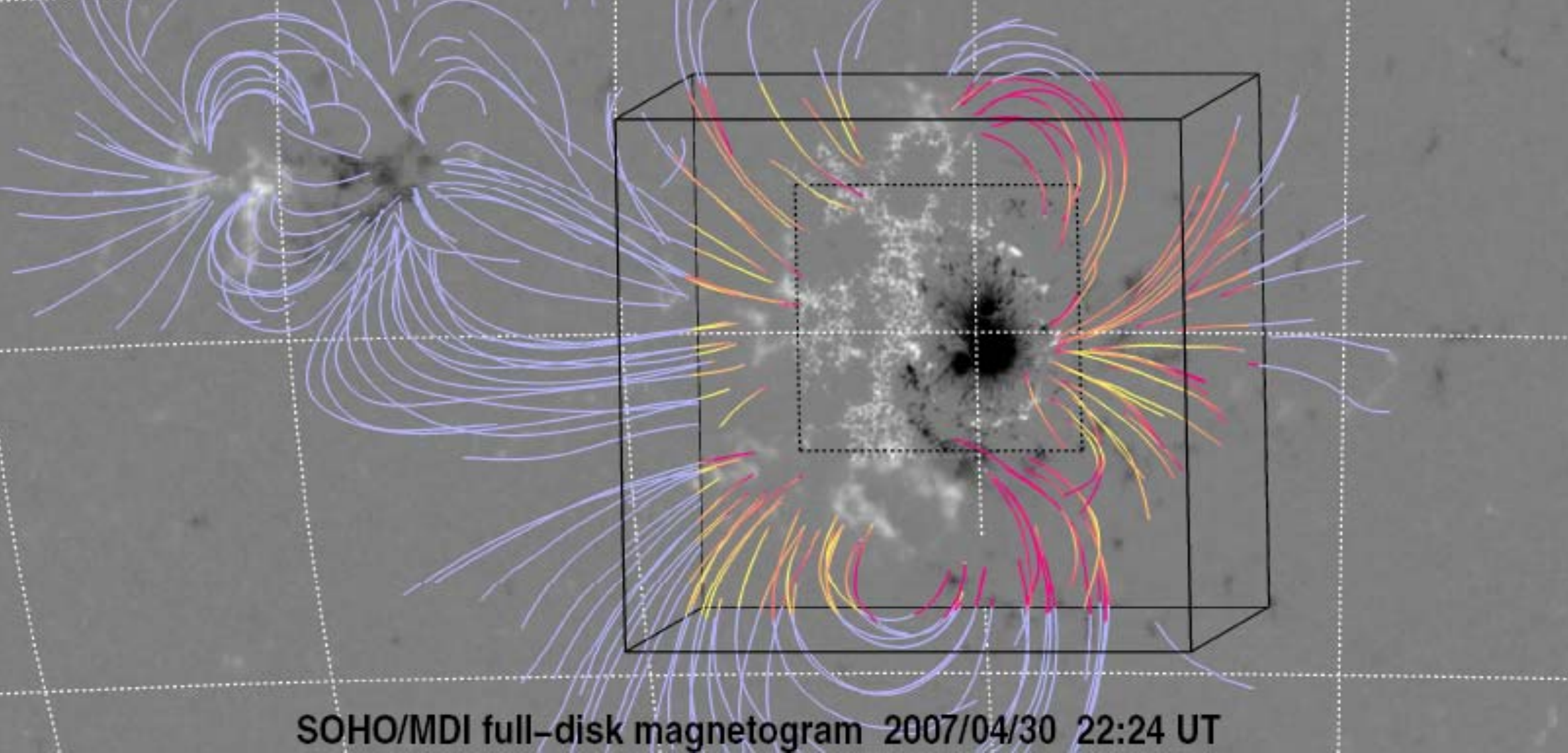}
\end{center}
\caption{SOHO/MDI magnetogram of AR 10953 with over-plotted stereoscopic
reconstructed loops \citep[see][]{aschwanden:etal08a} in blue and
extrapolated non-linear force-free coronal magnetic field lines.
The solid box depicts the $320 \time 320 \times 256$ grid point
simulation box in which the magnetic field extrapolations have
been carried out. The dotted area depicts the field-of-view of
Hinode/SOT-SP. Only in this dotted area photospheric vector
magnetograms have been available.
[Original published in \citep{derosa:etal09} as part of figure 1].}
\label{fig9}
\end{figure*}
As explained in the previous section, coronal magnetic field models
provide useful information for solving the stereoscopic
correspondence problem (see section \ref{sec2.2}) and to remove
reconstruction ambiguities (section \ref{sec2.4}). Until
now mainly linear force-free models have been used for this aim
and the method automatically fits the optimum linear
force-free parameter $\alpha$ for each loop individually
 { \refA to approximate its shape as closely as possible.
 $\alpha$ is here just used as a numerical parameter to alter the
 curve shape. It is therefore not surprising that
 \citep{feng:etal07a} found a significant scatter of $\alpha$ between
the loops. We cannot interpret the values of $\alpha$ in terms of a
linear force-free magnetic field model, because the different values are
a contradiction to the premises of
a globally constant $\alpha$. For a physical meaningful self-consistent
magnetic field model one cannot determine the values of $\alpha$ independently
for each loop.}
Unfortunately, a physical consistent
nonlinear model is  way more demanding, both computationally due to
the intrinsic nonlinearity of the model and from an observational
point of view since
nonlinear models require photospheric vector magnetograms as input.
Within the last few years a group of scientists (nonlinear force-free
field consortium, chaired by Carolus Schrijver) has intensively
compared and evaluated corresponding computer codes
\citep{schrijver:etal06,schrijver:etal08}, \citep{metcalf:etal08}, which
showed the codes produce reliable results, when feeded with consistent
input data (vector magnetograms or a quantity derived from vector magnetograms).
In another joint study of the consortium \citep[by][]{derosa:etal09} the codes
have been applied to AR 10953 and the 3D structure of the magnetic field
lines has been compared with the 3D geometry of plasma loops
\citep[as stereoscopically reconstructed in][]{aschwanden:etal08a}.
Figure \ref{fig9}), shows the stereo-loops in blue and in red and yellow
the magnetic field lines.
A major difficulty of this study was
that the Hinode-SOT/SP vector magnetograms, required as input for
the magnetic field codes, where available in only a very small
field of view (dotted area in figure \ref{fig9}) and
has only about $10\%$ of the area spanned by the stereo loops.
The majority of stereo-reconstructed loops were located outside of this
region, therefore the reconstruction
box was significantly enlarged beyond the Hinode-area
(solid lines in figure \ref{fig9}, but only the line of sight component
of the photospheric magnetic field from SOHO/MDI was available
in this enlarged box. Assumptions regarding the transverse photospheric
field component had to be made in the MDI-area, but unfortunately the
various magnetic field codes made different use of the assumption
in the MDI-area and consequently the resulting magnetic field lines
differed between the codes. The extrapolated magnetic field
lines turned out to be also inconsistent with the reconstructed STEREO-loops.
There was an average misalignment angle of $24^\circ$ between field lines and loops.
{\refA \cite{2009SoPh..tmp...81S} compared stereoscopic reconstructed
loops for three active regions observed in April and May 2007 with
potential field extrapolations and found a misalignment angle of
about $20^\circ-40^\circ$.}
In the \citet{derosa:etal09} study the nonlinear force-free approach did not provide
better agreement with the stereo-loops than the
much simpler to compute potential fields.
A main reason seems to be the very small field of view, where
actually vector magnetogram data have been available and correspondingly
the codes have not been fed with a consistent input. As visible
in figure \ref{fig9} almost no field line/STEREO-loop closes
within the Hinode field of view. It is therefore necessary to
repeat such a study with much larger field of view of
vector magnetogram data (ideally at least the same FOV as spanned
by the STEREO-loops).

{\refA An additional complication is that the magnetic field vector is
measured routinely only in the photosphere, where the magnetic field is
not force-free due to the high plasma $\beta$
\citep[see][]{metcalf:etal95}. Consequently photospheric
vector magnetograms do not provide consistent boundary conditions for
a nonlinear force-free extrapolation. To overcome this difficulty,
a preprocessing method has been developed to
remove the non-magnetic forces from the photospheric vector magnetograms
\citep[see][for details]{wiegelmann:etal06,fuhrmann:etal07}. These
preprocessed magnetograms are more chromospheric like and chromospheric
observations, e.g., can be incorporated into the preprocessing-algorithm
as described in \cite{wiegelmann:etal08}. In principle one could incorporate
additional observational constraints into the preprocessing routine, for
example minimize the angle of stereoscopic reconstructed loops (at the footpoints)
with the magnetic field vector. This might help to better estimate the magnetic
field vector in the upper chromosphere. Another possibility which might be
tried out is to add a term which minimizes the angle between the coronal magnetic field
and the reconstructed 3D coronal loops  in the nonlinear force-free
modelling algorithm. One possibility would be to extend the nonlinear force-free
optimization principle \cite{wheatland:etal00,wiegelmann04} by means of a
Lagrangian multiplier $\zeta$ as suggested in equation \ref{defLstereo}.}
\begin{eqnarray}
L&=&\int_{V} \left[B^{-2} \, |( \nabla \times {\bf B}) \times {\bf B}|^2 +|\nabla \cdot
{\bf B}|^2 \right] \; d^3V + \nonumber \\
&& \zeta \int_{V}({\bf B} \times {\bf S_{3D}})^2 \; d^3V
\label{defLstereo}
\end{eqnarray}
{\refA
The first two terms correspond to the force-free equations and the last term
measures the angle between the coronal magnetic field ${\bf B}$
and the stereoscopic reconstructed 3D-loops ${\bf S_{3D}}$, where ${\bf S_{3D}}$
should contain also an error approximation of the stereoscopic reconstruction error
This could be done by the local length of the vector ${\bf S_{3D}}$ along
the stereo-loop, where $|{\bf S_{3D}}|=1$ would indicate a small error and $|{\bf S_{3D}}|=0$
an infinite error. Locations with $|{\bf S_{3D}}|=0$ obviously do not contribute
to the functional. The third term in (\ref{defLstereo}) corresponds to a
weighted angle between magnetic field and STEREO-loops. Regions with high magnetic
field strength and accurate measurement of $|{\bf S_{3D}}|$ contribute more to the
functional.}
As such data (an active region observed
simultaneously  by both STEREO-spacecraft and  a large field-of-view
vector magnetograph) seem not to be currently available
(also due to a lack of active regions in recent years), corresponding studies
have to be postponed until after the launch of SDO.
The SDO/HMI intrument will provide full disc vectormagnetograph,
which resolves the limited FOV-problem of current instruments.
However, the angle between
the spacecraft might then be too large for stereoscopy with
STEREO A/B spacecraft.
One has to evaluate whether SDO/AIA and one of the STEREO-spacecraft can
be used for stereoscopy instead. The result can then be compared with nonlinear
force-free extrapolations from SDO/HMI. This instrument will provide the
required larger FOV for the field modelling.
%
\conclusions[Where to go in coronal stereoscopy?]
\label{sec4}
\begin{figure}
\vspace*{2mm}
\begin{center}
\includegraphics[width=8.3cm]{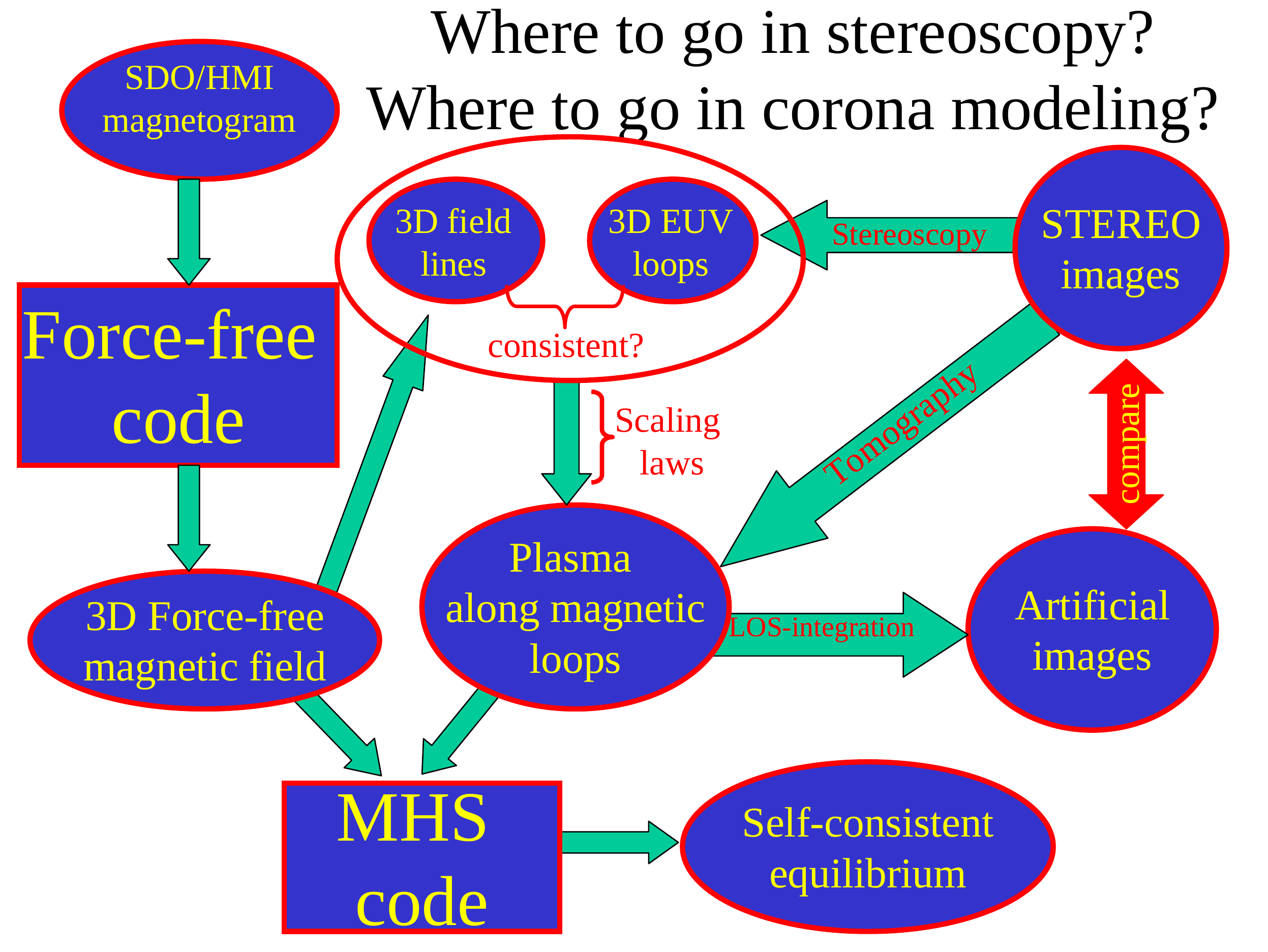}
\end{center}
\caption{A concept on how stereoscopy could be imbedded in a
self-consistent coronal modelling approach.}
\label{fig10}
\end{figure}
To summarize the current state of the art of coronal stereoscopy we propose
a concept of five steps, namely: feature extraction, association, geometric
reconstruction, error approximation and physical modelling.
 Several improvements are still possible
for some of the steps, e.g., a fully automatic  and reliable feature recognition
method, investigations on how stereoscopy is still possible with larger
separation angles between the spacecraft and if we can combine also STEREO
with other missions, e.g., SDO.
Some basic difficulties, e.g., the large reconstruction error on
the top of east-west loop is an intrinsic problem of stereoscopy, unless
we have one or more spacecraft well above or below the ecliptic.

Despite these principal difficulties coronal stereoscopy provides us
for the first time some information about the 3D geometry and
physical quantities of plasma
loops. A useful concept has been also to combine stereoscopy with
magnetic modelling of the solar corona, but due to shortage of
vector magnetogram data these approaches have been mainly done
with linear force-free methods. With the forthcoming full disc
vector magnetograph SDO/HMI nonlinear force-free magnetic field
models are assumed to become available on a regular basis.
A principal problem of force-free magnetic field modelling is
that it does not include a self-consistent modelling of the
coronal plasma. Assume a static corona
model for example, which balances the Lorentz-force
with pressure gradient and gravity as
$j \times B= \nabla p + \rho \nabla \psi$. The term force free means
that the Lorentz force vanishes and correspondingly the current
density $j$ is parallel to the magnetic field $B$. While this
approach is well justified for the field modelling in the low plasma
$\beta$ solar corona, it means also that
$ \nabla p + \rho \nabla \psi=0$, and because the gravity force
$-\nabla \psi$ is in the radial direction a consequence is that
the plasma would only be gravitationally stratified, contrary to observations.
{\refB
Consequently we cannot describe the solar coronal plasma self-consistently
within a force-free model.  Due to the low coronal plasma $\beta$ the
force-free assumption is well justified to compute the coronal magnetic
field structure. Variations in the plasma pressure along the field lines
are then compensated by relatively small Lorentz-forces.
For a self-consistent modelling
of the solar corona we have to take this effect into account and
use a magneto-hydro-static model in lowest order.}

In figure \ref{fig10} we outline an approach towards a self-consistent
coronal modelling. First there should be two alternative routes to
derive the 3D geometry of coronal loops, either from two EUV-images
by stereoscopy or by force-free extrapolations from vector magnetograms,
e.g., as provided in future from SDO/HMI. The resulting 3D field lines and
3D EUV-loops should be consistent with each other. Unfortunately this
has not been the case for a first comparison \citep[by][]{derosa:etal09}
as discussed already in section \ref{sec3}, but it is assumed
(or at least hoped for)that the new
instrumentation with SDO/HMI and improvements in modelling will lead to more
consistent results. If consistency between plasma loops and field lines has
been found, we will have a much more reliable magnetic field model
than we obtain from extrapolation alone.
One can then derive physical quantities along the loops as
outlined in section \ref{sec2.5}. An alternative method is to use
a tomographic approach \citep[as explained in detail in][]{aschwanden:etal09},
{\refA which provides the plasma density without model assumptions.
A sophisticated modelling of quantities along the loops can be done
\citep[see][who used scaling laws between the photospheric
magnetic field and the heating rate at the footpoints of loops]{schrijver:etal04}.}
From physical quantities like density and temperature one can compute
artificial EUV-images and compare them with the real STEREO-images.
This approach might as well allow to adjust free parameters within the
loop modelling approach. As pointed out above,
{\refB  modelling of coronal magnetic field and
plasma cannot be achieved a selfconsistently with a force-free model.
Plasma loop modelling
will create small gravity and plasma pressure forces, which have
to be compensated by a Lorentz-forces.
Due to the low coronal plasma $\beta$, the Lorentz force required
will be very small and the coronal magnetic field
structure will deviate only marginally from a force-free model.}

 Finally a self-consistent
model using the magneto-hydro-static approach can be computed with the
force-free magnetic field model and the plasma along the loops as input.
Corresponding magneto-hydro-static codes have been developed and
tested in \citep{wiegelmann:etal06b,wiegelmann:etal07} in cartesian and
spherical geometry, respectively. The resulting consistent static
model can be used as input to investigate dynamic phenomena for example
with time-dependent MHD-codes.

{
\refA
With stereoscopically reconstructed loops, even if they are few in number,
we will have for the first time an indirect though quantitative measurement of
the local coronal magnetic field direction. In future magnetic field extrapolation codes,
this information should be used not only for a comparison but as an additional constraint
to lift some of the uncertainties which arise from unknown boundary values. For a forecast
of the evolution and stability of observed active regions, we probably will need time
dependent simulations of the corona which incorporate observations by data assimilation
schemes similar to the way they are now used in computational meteorology
\citep[see][]{daley91}. An anfolding of the solar remote sensing observations
to constrain the 3D state of the solar corona is vital for these schemes and the
forecast range will strongly depend on the quality of the 3D reconstructions.
For this task, the techniques for detecting well defined objects in the images
and the way they are processed into 3D structures needs to be improved.
The present state of the art of solar stereoscopy can only be a first step in this direction.

A challenging task in stereoscopy is the 3D reconstruction of true
dynamic phenomena like the initiation of flares and CMEs. A key question is which
role magnetic reconnection plays for such eruptive phenomena. Stereoscopy could
help us first to derive the quasi-stationary state before an eruption, preferably
within a sophisticated self-consistent model, e.g., magneto-hydro-statics. If we
can then also observe the 3D-structure of the dynamic phase this would be very
helpful for our understanding of such phenomena. One could, in particular, use
the self-consistent stationary state as input for time dependent simulations,
e.g., with MHD or Hall-MHD. It is notoriously difficult to find transport
coefficients, e.g., resistivity, viscosity, heat tensor etc. for the coronal
plasma from micro-physics for the coronal plasma, because the required kinetic
scales are way to small to observe. One could, however, try to optimize these
transport coefficient with a systematic trial and error approach in order to
fit the observations best. Such observational based simulations would provide
us then a rich new world of insights. If the dynamic model simulations show
reasonable agreement with observed quantities, we might also have some confidence
in other model quantities, which cannot be observed. This could give insights about
the physics of magnetic reconnection in the coronal plasma.

A good knowledge of the physics of the solar corona,
say the 3D structure of loops and plasma along the loops will be very helpful,
together with high accuracy measurements of the photospheric magnetic field vector,
to understand the interface region between photosphere and corona. A physical
understanding of this region is important  for the coronal heating problem.
Coronal stereoscopy can help here because it provides a fair approximation of
the corona, but additional direct observations of the interface region, as
provided in near future for example from the small explorer mission IRIS are
necessary. A modelling approach in this region is very challenging, because
low and high $\beta$ plasma with sub- and supersonic plasma flows exist
side by side here.
}

%




\begin{acknowledgements}
We would like to thank the organization committee
(chaired by Richard Harrison) of the
STEREO-3/SOHO-22 workshop 'Three eyes on the Sun'
for the invitation to give this review on stereoscopy.
This work was supported by DLR-grant 50 OC 0501.
The SECCHI data
used here were produced by an international consortium of the
Naval Research Laboratory (USA), Lockheed Martin Solar and
Astrophysics Lab (USA), NASA Goddard Space Flight Center
(USA), Rutherford Appleton Laboratory (UK), University of
Birmingham (UK), Max-Planck-Institut for Solar System Research
(Germany), Centre Spatiale de Li\`{e}ge (Belgium), Institut
d'Optique Th\`{e}orique et Appliqu\`{e}e (France), and Institut
d'Astrophysique Spatiale (France).
\end{acknowledgements}
\bibliography{tw}
\bibliographystyle{copernicus}
\end{document}